\documentclass[11pt]{article}

\def\mytitle#1{\setcounter{equation}{0}
\setcounter{footnote}{0}
\begin{flushleft}\Large\textbf{#1}\end{flushleft}
\vspace{0.25cm}}
\def\myname#1{\leftline{{\large #1}}\vspace{-0.13cm}}
\def\myplace#1#2{\small\begin{flushleft}\textit{#1}\\
\texttt{#2}\end{flushleft}}

\usepackage{graphicx}
\begin{document}
\mytitle{Interacting Models of Generalised Chaplygin Gas and Modified Chaplygin Gas with Barotropic Fluid}

\myname{$Promila ~Biswas$\footnote{promilabiswas8@gmail.com} and $Ritabrata~
Biswas$\footnote{biswas.ritabrata@gmail.com}}
\myplace{Department of Mathematics, The University of Burdwan, Golapbag Academic Complex, City : Burdwan  - 713104, Dist. : Purba Barddhaman, State : West Bengal, India.} {}
 
\begin{abstract}
In this letter we consider two different models of our present universe. We choose the models which are consisting different sets of two seperate fluids. The first one of each set tries to justify the late time acceleration and the second one is  barotropic fluid. The former model considers our present time universe to be homogeneously filled up by Generalized Chaplygin Gas which is interacting with barotropic fluid. On the other hand, the latter model considers that the cosmic acceleration is generated by Modified Chaplygin Gas which is interacting with matter depicted by barotropic equation of state. For both the models, we consider the interaction term to vary proportionally with Hubble's parameter as well as with the exotic matter/dark energy's energy density. We find an explicit function form of the energy density of the cosmos which is found to depend on different cosmological parameters like scale factor, dark energy and barotropic fluid's EoS parameters and other constants like interacting constants etc. We draw curves of effective EoS-s, different cosmological parameters like deceleration parameter $q$, statefinder parameters $r$ and $s$ with repect to the redshift $z$ (for different values of dark energy and barotopic fluid parameters) and study them thoroughly. We compare two models as well as the nature of dependencies on these models' interaction coefficients. We point out the particular redshift for which the universe may transit from a deceleration to acceleration phase. We tally all these values with different observational data. Here we also analyse how this value of particular redshift does change for different values of interaction coefficients and different dark energy models.
\end{abstract}
\section{Introduction}
Recent contents of redshift and luminosity-distance relations of highly redshifted type Ia Supernovae ($SNeIa$) \cite{Observational_Evidence_from_Supernovae, Measurements_of_Omega, Type_Ia_Supernova_Discoveries} specify that the expansion of the universe is accelerating \cite{Observational_Evidence_from_Supernovae, 
Measurements_of_Omega, Type_Ia_Supernova_Discoveries, Cosmology_from_Type_Ia_Supernovae, Constraints_on_Cosmological_Models}. To support such observations of late time acceleration theoretically, we need to modify existing theory of gravity and/or stress energy part which is present in the space time. A part of such studies speculates that the current phase of our universe is possibly dominated by a smooth energy component exerting negative pressure. To obtain the cosmic acceleration, directed by the Friedmann Lemaitre Robertson Walker (FLRW) metric the pressure $p$ and the energy density $\rho$ of the universe should violate the strong energy condition $3p + \rho > 0$ and then the weak energy condition $p+\rho>0$. The imaginary stress energy responsible for such cosmic acceleration is popularly referred to as Dark Energy (DE here after) \cite{The_Case_for_a_Positive_Cosmological_Lambda, The_Cosmological_Constant_and_Dark Energy,  Cosmological_Constant_the_Weight_of_the_Vacuum}. 

In addition with the $SNeIa$ observations, measurements of the cosmic microwave background (CMB) \cite{First_Year_Wilkinson_Microwave_Anisotropy_Probe} as well as that of the galaxy power spectrum also indicate the existence of the DE. From the Wilkinson Microwave Anisotropy Probe (WMAP) satellite observation, we have come to know that DE, dark matter (DM hereafter) and the usual baryonic matter occupy respectively about $73\%$, $23\%$ and $4\%$ of the total mass-energy budget of the present time universe. There exist different proposed candidates to play the role of DE. The most traditional candidate is a nonvanishing cosmological constant $\Lambda=-1$ \cite{de_Sitter_Vacua_in_String_Theory} which can also be thought of as a perfect fluid satisfying the equation of state(EoS hereafter) $p_{\Lambda} = -\rho_{\Lambda}$ and it suffers from conceptual problems such as fine-tuning problem and coincidence problem \cite{The_Cosmological_Constant_Problems}. A number of viable models for DE also
have been constructed. Some of the scenario are quintessence \cite{Cosmology_and_the_Fate _of_Dilatation_Symmetry, Cosmological_consequences_of_a_rolling_homogeneous_scalar_field}, Chameleon \cite{Chameleon_Fields:_Awaiting_Surprises_for_Tests_of_Gravity_in_Space}, K-essence \cite{Kinetically_driven_quintessence, A_Dynamical_Solution} ( based on earlier work of K-inflation \cite{k_Inflation}), modified gravity \cite{4D_Gravity_on_a_Brane_in_5D_Minkowski_Space, Introduction_to_Modified Gravity}, Tachyon \cite{Accelerated_expansion_of_tachyonic_matter} arising in string theory \cite{Tachyon_Matter}, quintessential inflation \cite{Quintessential_inflation} and Chaplygin gas (CG hereafter) (which attempts to unify DE and DM under one system by an EoS \cite{An_alternative_to_quintessence} $p_d = -\frac{B}{\rho_d}$, $B > 0$ ($\rho_d$ and $p_d$ are the DE density and DE pressure respectively) which evolves between the two exotic fluids \cite{An_alternative_to_quintessence, Generalized _Chaplygin_Gas} and the generalized Chaplygin gas (GCG hereafter) \cite{Unification_of_Dark_Matter, Exploring_the_Expanding_Universe} with the exotic EoS : 
\begin{equation}\label{GCG}
p_d = -\frac{B}{\rho_d^\alpha}~~,~~0 < \alpha \leq 1
\end{equation}
Negative pressure leading to an accelerating universe can be obtained in CG cosmology \cite{An_alternative_to_quintessence}. CG behaves as pressureless fluid for small values of the scale factor. It behaves as a cosmological constant for large values of the scale factor which tends to accelerate the present time expansion.  The value of $A$ is constrained in the region $A=0.87^{+0.13}_{-0.18}$\cite{data_for_A}. $\alpha=0.2$ is a favourable value for galaxy clusters $X$-ray and supernova data \cite{data_for_alpha_1}. Analysis of the reference \cite{data_for_alpha_2} predicts the value of $\alpha$ to be nearly equal to $1.18^{+4.12}_{-2.18}$ for flat universe. Also, the GCG model can be extended to the Modified Chaplygin Gas (MCG) model \cite{Accelerated_Universe_from_Modified_Chaplygin_Gas, The_modified_Chaplygin_gas, A_modified_Chaplygin_gas_model_with_interaction} :
\begin{equation}\label{MCG}
p_d = A\rho_d - \frac{B}{\rho_d^\alpha}~~,
\end{equation}
where $A$ and $B$ are positive constants and $0 < \alpha < 1$, which interpolates between standard fluid at high energy densities and CG fluid at low energy densities and can also describe the current accelerating expansion of the universe. Various other modifications of CG have appeared in the literature such as variable CG, holographic and interacting holographic CG, viscous CG models etc amongst others. However, each one of them comes with both merits and demerits as far as related cosmology is concerned. Generalized Cosmic Chaplygin Gas \cite{You_Need_Not_Be_Afraid, Role_of_Generalized_Cosmic_Chaplygin_Gas_in_Accelerating_Universe} (GCCG) was introduced in 2003 which can be made to be stable and free from unphysical behaviours even when the vacuum fluid satisfies the phantom energy condition, which is the striking factor of this model. 

In this letter, a flat Friedmann-Lemaitre-Robertson-Walker (FLRW) universe has been considered and assumed to be filled up with two fluids, i.e., GCG or MCG and a barotropic fluid with EoS $p_m = w_m\rho_m$, $w_m\geq -\frac{1}{3}$ is taken to be a constant. The lower bound on $w_m$ assures that the barotropic fluid does not violate the strong energy condition. 

Energy densities of radiation($r$), baryons($b$), cold dark matter($c$) and dark energy($d$) are taken to be conserved seperately in concordance model : this implies the energy conservation equations as 
\begin{equation}
\dot{\rho_i}+ 3H(\rho_i+p_i) = 0~~~~i=r,~b,~c,~d.
\end{equation}
For interacting DM-DE models, the total energy density of the dark sector is conserved. Seperately the densities of DE and DM evolve as
\begin{equation}
\dot{\rho}_c+3H\rho_c=Q~~and~~\dot{\rho}_d+3H(1+w_d)\rho_d=-Q~~,
\end{equation}
where $Q$ signifies the interaction kernel. In particle physics, a researcher can find the kernel to be a function of the energy densities ($\rho_c$, $\rho_d$) involved and of time ($H^{-1}$). The first order Taylor expansion of such interaction will be $Q=H\left({\cal C}_d\rho_d+{\cal C}_c\rho_c\right)$ with ${\cal C}_d$ and ${\cal C}_c$ as constants to be determined observationally\cite{r1}. A Bayesian evidence that the strength of the coupling with CDM varies with redshift is studied in the reference \cite{r2}. Testing of coupled DE models with their cosmological background evolution is done in the reference \cite{r3}. In the reference \cite{r4}, it has been considered that the energy transfer rate was proportional to the Hubble parameter and energy density of DE $Q=3Hc_i \rho_i$ or $Q$ is found to be taken as a linear combination of products of Hubble's parameter and energy density(ies) of DE or/and DM. Such a system is observationally supported/ constrained in a physical domain in the reference \cite{r5}. Constraining different parameters of this kind of interaction can be found in the references\cite{r6, r7, r8}. The authors of the reference \cite{CG_revisited} have already studied about the interaction between CG and Barotropic Fluid.

Our investigation has been primarily focussed on the dynamics of the coexistence of the fluids in the presence of an interaction term proportional to the Hubble parameter times the DE density. This class of interaction terms generally appears in the interacting holographic dark energy (HDE) model. In the absence of interaction, there exists no scaling solutions owing to the fact that the EoS of MCG decreases with scale factor while the DM EoS remains constant. Moreover, their effective EoS could also cross the phantom barrier. However, their form of interaction failed to produce an analytic solution which is necessary to obtain in order to have a clear and a nice picture of the cosmological model concerned. We wish to observe the variation of dimensionless density parameters of different matter content of the universe and their dependencies on different interaction parameters as well as the EoS parameters of DE. Besides the nature of deceleration parameter, values of redshift for which deceleration to acceleration take place, natures of statefinder parameters will be studied and we will try to find the dependencies of these parameters on the interaction and DE EoS parameters.

Our letter is organized as follows : in section 2 we describe the basic equations that govern a flat FLRW universe filled with GCG or MCG and barotropic fluid. Next, it is concerned with the cosmological implications of considering interactions between GCG and barotropic fluid. Then, we describe the similar way of interactions between MCG and barotropic fluid and the corresponding graphs are also plotted for this case. Finally, we  will conclude our letter.
\section{Mathematical Formulation of the Problem}
We consider a flat Friedmann-Lemaitre-Robertson-Walker universe which is governed by the metric[$c=1$]
\begin{equation}
ds^2=-dt^2+a^2(t) \Big[ dr^2+r^2(d\theta^2+sin^2\theta d\phi^2) \Big],
\end{equation}
where $a(t)$ is the scale factor of the universe. We assume a perfect fluid is filled up all over in the universe having
energy-momentum tensor :
\begin{equation}
T_{\mu\nu} = (\rho + p)u_\mu u_\nu + pg_{\mu\nu} ~~~~,
\end{equation}
$u^{\mu}=\frac{dx^{\mu}}{d \tau} $ is the 4-velocity of the fluid, $\rho$ is the total energy density of the universe and $p$ is the pressure term. Here, we have assumed that $ 8\pi G = 1$. The Friedmann equations can be obtained as
\begin{equation}\label{field_equation_I}
H^2 = \frac{1}{3}\rho
\end{equation}
and
\begin{equation}\label{field_equation_II}
\dot{H} = -\frac{1}{2}(\rho + p)
\end{equation}
where $H$ is the Hubble parameter defined by $H=\frac{\dot{a}(t)}{a(t)}$.

Since we shall be working with a two-fluid system at a time, i.e., first we will show the interactions between GCG and barotropic fluid and then MCG and barotropic fluid. The total energy density $\rho$ and the total pressure $p$ can be written as 
\begin{equation}\label{density_equation}
\rho = \rho_m + \rho_d
\end{equation}
and
\begin{equation}\label{pressure_equation}
p = p_m + p_d~~~~~~~~~.
\end{equation}
Using (\ref{density_equation}) and (\ref{pressure_equation}), we get from (\ref{field_equation_I}) and (\ref{field_equation_II}), the Friedmann equations as
\begin{equation}
H^2 = \frac{1}{3}(\rho_m + \rho_d)
\end{equation}
and
\begin{equation}
\dot{H} = -\frac{1}{2}(\rho_m + \rho_d + p_m + p_d)~~~~~~~~~.
\end{equation}

Interactions between DE and DM have some important consequences such as in alleviating the coincidence problem \cite{Coupled_Quintessence, Interacting_quintessence_solution, Observational_constraints_on_interacting_quintessence_models, 
Coupled_quintessence_and_the_coincidence_problem, Interacting_Dark_Matter_and_Dark_Energy, Holographic_dark_energy_and_cosmic_coincidence}, among others. The coincidence problem can be solved if DE decays into DM \cite{Interacting_Quintessence}, thus reducing the difference between the densities of the two components through the evolution of the universe. Then the interaction term between the GCG and the barotropic fluid is taken as the following form \cite{Interacting_quintessence_solution, Interacting_Quintessence}
$Q = 3b^2H\rho_d$. Therefore, if GCG is assumed to decay into matter, then under the above form of interaction, the conservation equations are as follows,
\begin{equation}\label{energy_density}
\dot{\rho_d} + 3H(\rho_d+ p_d) = -Q = -3b^2H \rho_d
\end{equation}
and
\begin{equation}\label{Matter_density}
\dot{\rho_m} + 3H(\rho_m + p_m) = +Q = +3b^2H\rho_d~~~~~~~~~. 
\end{equation}
As $p_m = w_m\rho_m$, from (\ref{Matter_density}) we get,
\begin{equation}
\dot{\rho_m} + 3H(1 + w_m)\rho_m = +Q = +3b^2H\rho_d 
\end{equation}
\subsection{Generalised Chaplygin Gas With Barotropic Fluid}
From (\ref{energy_density}), for GCG we get, the continuityequation as
$$ \dot{\rho_d} + \frac{3 \dot{a}}{a}\bigg[\rho_d + b^2 \rho_d - \frac{B}{\rho_d^{\alpha}}\bigg] = 0\Rightarrow \large \int \frac{\rho_d^\alpha~~ d\rho_d}{(1+b^2)\rho_d^{1+\alpha}-B}= - \large \int \frac{3}{a}da~~~~.$$
Integrating we get,
\begin{equation}\label{rho_d_for_gcg}
\frac{1}{(1+b^2)(1+\alpha)} ln \left| (1+b^2)\rho_d^{1+\alpha}-B \right| = ln B' - ln a^3\Rightarrow \rho_d = \frac{1}{(1+b^2)^{\frac{1}{1+ \alpha}}}\left[B+ \left(\frac{B'}{a^3}\right)^{(1+b^2)(1+\alpha)}\right]^\frac{1}{1+\alpha}~~,
\end{equation}
where $B'$ is integrating constant. To make the expression easier we will write $C=B'^{(1+ b^2)(1+\alpha)}$. Now, using the value of  $\rho_d$ and equation (\ref{rho_d_for_gcg}) in equation (\ref{Matter_density}) and multiplying both sides of the equation by $a^{3(1+w_m)}$, $\rho_m$ can be evaluated as
$$a^{3(1+w_m)} \rho_m = \frac{3 b^2}{(1 + b^2)^\frac{1}{1 + \alpha}} \int a^{2(1+w_m)} \times \bigg[ \{ B + \frac{C}{a^{3(1+b^2)(1+\alpha)}} \}-(1+w_m)\bigg] da$$
$$\Rightarrow \rho_m = \frac{1}{a^{3(1+w_m)}} \Bigg[C'+ a^{3(1+w_m)} \bigg\{ -1 + \frac{b^2(1+b^2)^{-\frac{1}{1+\alpha}}}{1+w_m}\bigg(B + \frac{C}{a^{3(1+b^2)(1+\alpha)}} \bigg)^\frac{1}{1+\alpha}$$
\begin{equation}
\times \bigg(1+\frac{C}{Ba^{3(1+b^2)(1+\alpha)}}\bigg)^{-\frac{1}{1+\alpha}} \times _2F_1 \bigg[ -\frac{1}{1+\alpha}, -\frac{1+w_m}{(1+b^2)(1+\alpha)}, 1- \frac{1+w_m}{(1+\alpha)(1+b^2)}, \frac{C}{a^{3(1+b^2)(1+\alpha)}}\bigg]\bigg\}\Bigg]~~~~,
\end{equation}
where $C'$ is integrating constant.  $_2F_1[y_1, y_2, y_3, x]$ is known as Gauss's hypergeometric function \cite{Hypergeometric_function}. From (\ref{density_equation}) we can derive the expression for total density as, 
$$\rho =  \frac{1}{a^{3(1+w_m)}} \Bigg[C'+ a^{3(1+w_m)} \bigg\{ -1 + \frac{b^2(1+b^2)^{-\frac{1}{1+\alpha}}}{1+w_m}\bigg(B + \frac{C}{a^{3(1+b^2)(1+\alpha)}} \bigg)^\frac{1}{1+\alpha}\times \bigg(1+\frac{C}{Ba^{3(1+b^2)(1+\alpha)}}\bigg)^{-\frac{1}{1+\alpha}}$$
\begin{equation}\label{16}
 \times _2F_1 \bigg[ -\frac{1}{1+\alpha}, -\frac{1+w_m}{(1+b^2)(1+\alpha)}, 1- \frac{1+w_m}{(1+\alpha)(1+b^2)}, \frac{C}{a^{3(1+b^2)(1+\alpha)}}\bigg]\bigg\}\Bigg]+ \frac{1}{(1+b^2)^{\frac{1}{1+ \alpha}}}\left[B+ \frac{C}{a^{3(1+b^2)(1+\alpha)}}\right]^\frac{1}{1+\alpha}
\end{equation}
and from (\ref{pressure_equation}) and (\ref{16}) we can evaluate the expression for GCG pressure as
$$p = \frac{w_m}{a^{3(1+w_m)}} \Bigg[C'+ a^{3(1+w_m)} \bigg\{ -1 + \frac{b^2(1+b^2)^{-\frac{1}{1+\alpha}}}{1+w_m}\bigg(B + \frac{C}{a^{3(1+b^2)(1+\alpha)}} \bigg)^\frac{1}{1+\alpha}\times \bigg(1+\frac{C}{Ba^{3(1+b^2)(1+\alpha)}}\bigg)^{-\frac{1}{1+\alpha}} \times$$
$$ _2F_1 \bigg[ -\frac{1}{1+\alpha}, -\frac{1+w_m}{(1+b^2)(1+\alpha)}, 1- \frac{1+w_m}{(1+\alpha)(1+b^2)}, \frac{C}{a^{3(1+b^2)(1+\alpha)}}\bigg]\bigg\}\Bigg]- B \bigg[(1+b^2)^{\frac{\alpha}{1+ \alpha}}\left\{B+ \frac{C}{a^{3(1+b^2)(1+\alpha)}}\right\}^{-\frac{\alpha}{1+\alpha}}\bigg] $$
$$~~~~~~~~.$$
The variations of the DE density $w_d = - \frac{B}{\rho_d^{1 + \alpha}}$, the effective EoS parameter $w_{eff} = \frac{p}{\rho}$ and the deceleration parameter $q = \frac{3}{2}(1 + \frac{p}{\rho})-1$ for this interacting scenario are now constructed and graphically analysed.
\begin{figure}[!ht]
\begin{center}
\includegraphics[height=9in, width=7in]{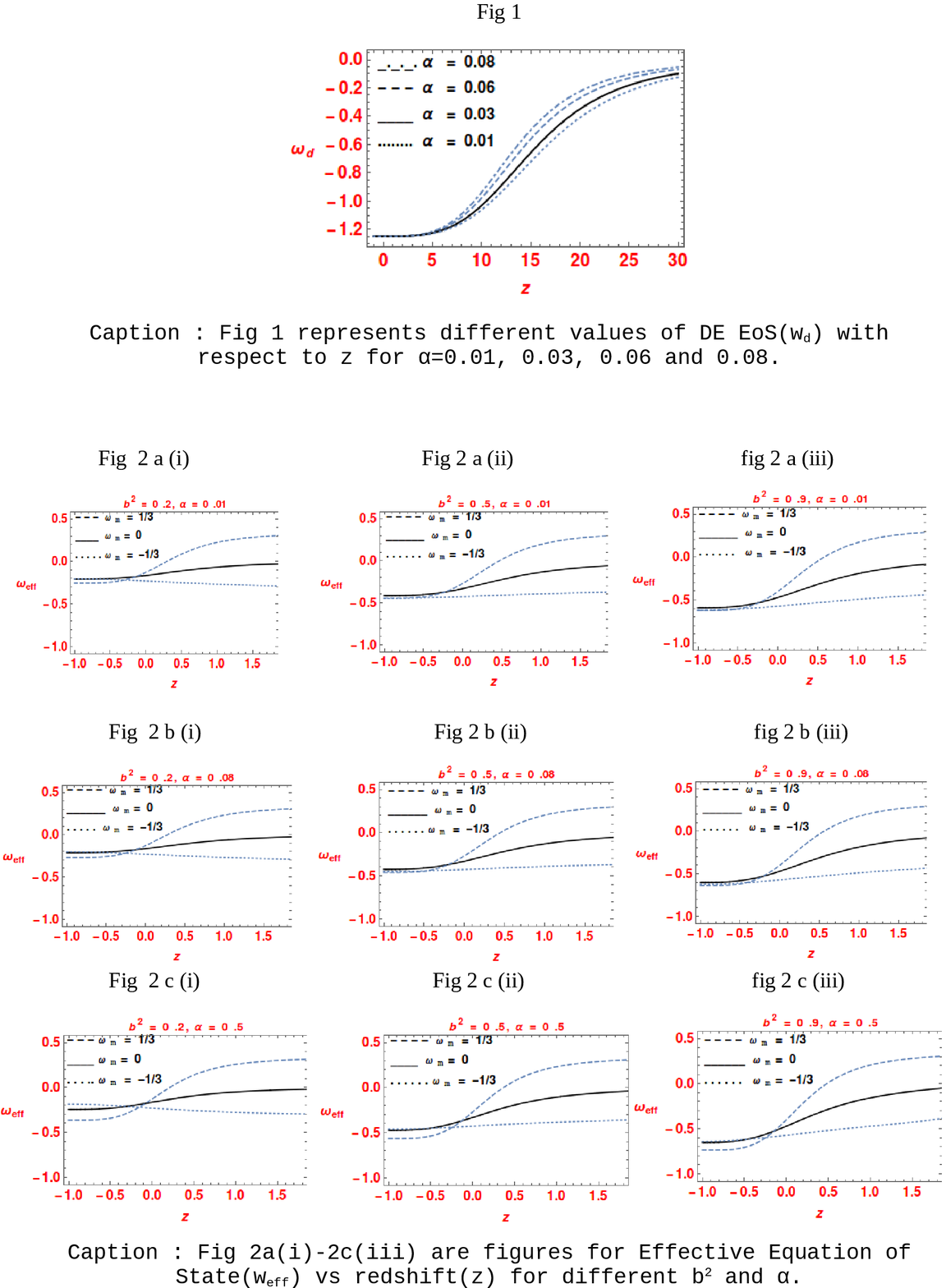}~~\\
\end{center}
\end{figure}
We plot $w_d$ vs $z$ in figure 1. We observe, for a particular $z$, $w_d(\alpha_1)>w_d(\alpha_2)$ if $\alpha_1>\alpha_2$. For low $z$, all the curves become parallel to $z$-axis and as we increase $z$, the differences between the values of $w_d$ for different $\alpha$-s become clearer. This means, in the neighbourhood of present time, GCG's $w_d$-s are almost equal and very much negative. In past we can have different negative pressures depending on different values of $\alpha$. As we go to the $``paster"$ region (i.e., where $z$ is high), values of $w_d$-s become almost constant (but negative).

We will vary $w_{eff}$ with respect to $z$ in figures 2a(i) to 2c(iii). Row wise we will increase the values of $b^2$ and column wise we will increase the values of $\alpha$. It is clear from the comparison of 2a(i) and 2c(iii) that high $(b^2,~\alpha)$ case increases the rate of increment of $w_{eff}$ for $w_m=0$. For $w_m=\frac{1}{3}$, i.e., radiation interacted with DE, the graph shows lesser value of $w_{eff}$ in future $(z<0)$ and higher in past $(z>0)$. When high value of $(b^2,~\alpha)$ is considered, for $w_m=-\frac{1}{3}$ we see the $w_{eff}$ to decrease with time. For low $b^2$, this decreasing tendency is carried over. But if we increase $\alpha$, $w_{eff}$ for $w_m=-\frac{1}{3}$ case increases. This signifies if radiation interacts with DE, the effective EoS reduces. But as we increase the interacting parameter $b^2$, the effective EoS starts to increase.
\begin{figure}[!ht]
\begin{center}
\includegraphics[height=9in, width=7in]{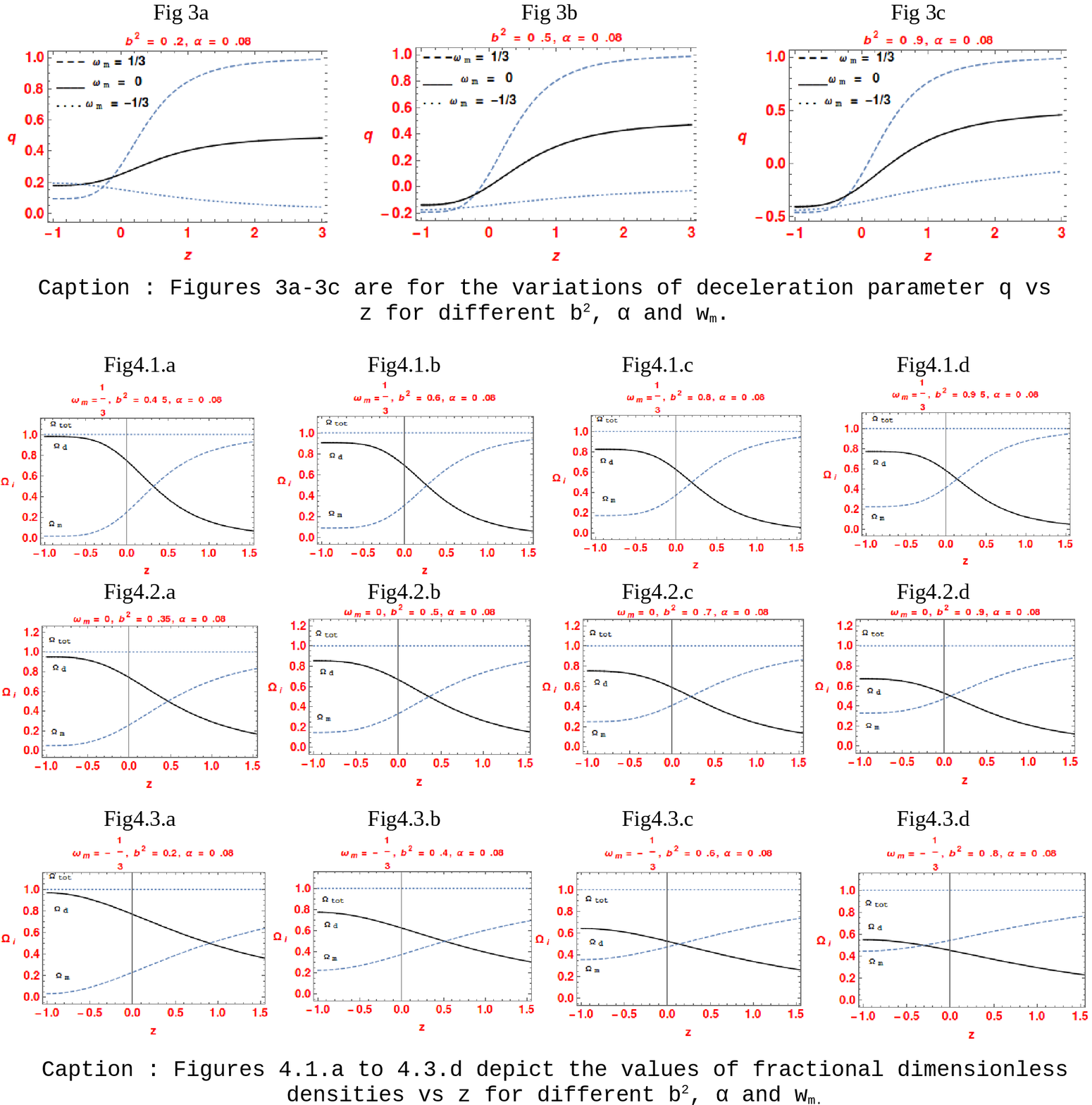}~~~
\end{center}
\end{figure}
For $w = \frac{1}{3}$, i.e., radiation case, we observe that the $w_{eff}$ is positive but constant for positive $z$ and at present time's neighbourhood. It starts to reduce and in future and it becomes negative but constant. This future $w_{eff}$ constant line is however stays higher than the $w_{eff} = -1$ line. This signifies that the $w_{eff}$ is  reaching near to the phantom barrier in future.

For pressureless dust($w_m=0$) working together with GCG, we see a slowly decreasing line with time. This is very small and positive in past whereas small but negative in future. For $w_m = -\frac{1}{3}$, i.e., for a matter exerting negative pressure (particularly, exotic matter at quintessence barrier) we observe just the opposite. The past shows a negative $w_{eff}$ whereas the future $w_{eff}$ is positive saying that if GCG is working with a negative pressure exerting exotic matter, it is not at all negative pressure exerting anymore. 

The deceleration parameter has almost the same nature as the $w_{eff}$ has. This is plotted for GCG in Fig 3a-3c. As we increase the value of $b^2$, we observe that the span of $q$ for both $``less~ z"$ or $``high ~z"$ reduces. For $w_m=-\frac{1}{3}$ case, the deceleration parameter reduces if $b^2$ is low. For high $b^2$, $q$ is increasing. For $b^2=0.2$, all the deceleration parameter values are positive. But for high $b^2$ values we observe deceleration parameter to enter in the positive range from negative range.

Using the same set of values for the free parameters, we have plotted the variation of the fractional energy densities of DE, $\Omega_d = \frac{\rho_d}{\rho}$ and that of matter, $\Omega_m = \frac{\rho_m}{\rho}$ respectively and presented them in Figure 4.1.a - 4.3.d.

Fractional dimensionless densities should lie in the envelope $0 < \Omega_i \leq 1$. For interaction of GCG and radiation if $b^2$ is taken to be $0.45$, $\Omega_m$ is low (asymptotic to $\Omega_i = 0$ line) for negative $z$ (i.e., for future) and high (asymptotic to $\Omega_i = 1$ line) for high $z$ (i.e., in past). $\Omega_d$ has just the opposite property. $\Omega_d$ and $\Omega_m$ do have the same values at some $z_{d=m}(b^2)$. Physically this conveys us that the fractional density of DE was low in past and high in future and converse for radiation density. But as we increase $b^2$, we see the maximum value to be attained by $\Omega_d$ in future goes down and the minimum value to be achieved by $\Omega_m$ increases, $z_{d=m}(b_1^2) ~>~ z_{d=m}(b_2^2)$ if $b_1^2 > b_2^2$.

As we change the scenario to the interaction between pressureless dust and GCG, we observe the highest value achieved by $\Omega_m$ in past decreases and the lowest value occured by $\Omega_d$ in past increases. $z_{d=m}(b_1^2)$ for $\omega_m = \frac{1}{3} <~$ that for $\omega_m = 0$. This same trend takes place if we take the interaction between quintessence with GCG. For high $b^2$, in future region we observe these two fluids to have nearby fractional densities.

If the EoS of the fluid which is going to interact with GCG decreases, we fluid their fractional densities to have same values at ``paster" regions. Increment in $b^2$ leads these fractional densities to have same values in ``futurer" regions. Increment in $b^2$ forces DE to have lesser fractional densities.

The models are trying to provide a description of the cosmic acceleration are proliferating, there exists the problem of discriminating between the various contenders. To this aim Sahni et al \cite{Statefinder_a_new_geometrical} proposed a pair of parameters $\{ r, s\}$ called statefinder parameters. In fact, trajectories in the $\{ r, s\}$ plane corresponding to different cosmological models demonstrate qualitatively different behaviour. The statefinder parameters are defined via the
following pairs of parameters 
\begin{equation}
r \equiv \frac{\frac{d^3a}{dt^3}}{aH^3}~~and
\end{equation}   
\begin{equation}
s \equiv \frac{(r-1)}{3(q-\frac{1}{2})}~~.
\end{equation}
The new feature of the statefinders are that they involve the third derivative of the cosmological radius. These parameters are dimensionless and allow us to characterize the properties of DE. Trajectories in the $(s - r)$-plane corresponding to different cosmological models, for example $\Lambda$CDM model's diagrams correspond to the fixed point $s = 0$, $r = 1$.

We have also calculated the redshift of transition from deceleration to acceleration ($z_{da}$). But we have noticed that there is no $z_{da}$ value for $b^2 = 0.2$.
\\

\textbf{TABLE I} : {\bf Values of the cosmological parameters with $B = 5,~\alpha = 0.08,~ b^2 = 0.5,~ C = 0.0001,~ C' = 1$.}
\begin{center}
 \begin{tabular}{|| c | c | c | c ||}
 \hline
 Parameter & $w_m = \frac{1}{3}$ & $w_m = 0$ & $w_m = -\frac{1}{3}$ \\ [0.5ex]
 \hline
$z_{da}$ & -0.11299 & 0 & 5.04565 \\
 \hline
  $q$ &  0.0000427558 &0.0000219354 & 0.0000821006 \\
 \hline
 $\Omega_d$ & 0.799982 & 0.666677 & 0.0828642  \\
 \hline
 $\Omega_m$ & 0.200018 & 0.333323 & 0.917136  \\
  \hline
  $r$ & 0.0000427594 & 0.0000219364 & 0.0000821141 \\
  \hline
  $s$ & 0.666695 & 0.666681 & 0.666721 \\ [1ex]
  \hline
\end{tabular}
\end{center}

\textbf{TABLE II} : {\bf Values of the cosmological parameters with $B = 5,~\alpha = 0.08,~ b^2 = 0.9,~ C = 0.0001,~ C' = 1$.}\\
\begin{center}
 \begin{tabular}{|| c | c | c | c ||}
 \hline
 Parameter & $w_m = \frac{1}{3}$ & $w_m = 0$ & $w_m = -\frac{1}{3}$ \\ [0.5ex]
 \hline
$z_{da}$ & 0.093279 & 0.43555 & 4.0465 \\
 \hline
  $q$ &  0.0000102566 & 0.0000213814 & 0.0000381052 \\
 \hline
 $\Omega_d$ & 0.540564 & 0.370501 & 0.111312  \\
 \hline
 $\Omega_m$ & 0.459436 & 0.629499 & 0.888688  \\
  \hline
  $r$ & 0.0000102569 & 0.0000185957 & 0.0000381081 \\
  \hline
  $s$ & 0.666674 & 0.666679 & 0.666692 \\ [1ex]
  \hline
\end{tabular}
\end{center}
For observation supported values of $q$, we see, there exist plenty of articles. The author of the reference \cite{MNRAS_422_2012} has used $H(z)-z$ data and SNeIa data. The deceleration parameter is constrained for a power law expansion. At $1\sigma$ level, the constraint from $H(z)$ data\footnote{Simon, Verde and Jimenez \cite{Constraints_on_the_redshift_dependence} determined nine $H(z)$ data points in the range $0 \leq z \leq 1.8$ by using the differential ages of passively evolving galaxies determined from the Gemini deep deep survey and archival data. Another set of $H(z)$ data at eleven different red shifts based on the differential ages of red-envelope galaxies were reported by Stern et. al \cite{Stern_Jimenez} where three mere $H(z)$ data points were obtained by Gaztanaga, Cabre, and Hui \cite{Clustering_of_luminous}.} is obtained as $q = -0.18^{+0.02}_{-0.12}$ while the constraint from type Ia supernovae data\footnote{Union two set of 557 $SNeIa$ from supernovae cosmology is used \cite{Spectra_and_Light_Curves}} is $q = -0.38^{+0.08}_{-0.05}$. The joint test using $H(z)$ and SNeIa data yields the constraints $q = -0.34^{+0.05}_{-0.05}$. The tests of power law cosmology accommodates well the $H(z)$ and $SNeIa$ data using the primordial nucleosynthesis which yields the constraints $q$ greater than or approximately equal  to  $0.72$. We find some other literature \cite{Cunha1} where the transition redshift  is found to be $0.49^{0.135}_{-0.07}(1\sigma)^{+0.54}_{-0.12}(2\sigma)$. Cosmographic analysis of present time deceleration parameter is done in the reference \cite{Gruber1, Aviles1}. Comparative studies of effective EoS parameter are done in \cite{Singh1}. This work discusses about the cosmological parameters for different cosmic times. Scale factor -cosmic time relations are analysed.

For our analysis, we observe that the value of $q$ for $w_m = \frac{1}{3}$ is almost supported by SNeIa and combined $H(z)$- SNeIa data if $z_{{da}_1}$ is concerned. But for $w_m = 0$, the value of $q$ derived by us is approximately supported by only $H(z)$ data. Table I and Table II consist different values of different cosmological parameters such as deceleration parameter, dimensionless densities of DE and the interacting barotropic fluid, statefinder parameters $r$ and $s$ for the particular redshift $z_{da}$ where deceleration to acceleration takes place. The occurance of such $z_{da}$ depends on the nature of the interacting barotropic fluid, i.e., on the EoS $w_m$ of the fluid, interacting DE's EoS parameter $B$ and $\alpha$, interaction coefficient $b^2$ and on some other constants of integration which arise from the integration of continuity equation for DM and interacting matter. As we decrease the barotropic matter's EoS from positive to negative via zero, the value of redshift to transit from deceleration to acceleration increases. This denotes that deceleration to acceleration occurs in future if the matter part is radiation like whereas the tansition from deceleration to acceleration has been ocurred in past if the interacting barotropic fluid behaves as a negative pressure exerting fluid (which is obvious as it is easy for a negative pressure exerting matter to initiate acceleration when it is working together with a DE model). If we increase the value of $b^2$, i.e, force DE to convert into matter, we see the acceleration to get initiated in past. Dimensionless density for DE decreases with increment of $b^2$ at $z_{da}$ whereas the opposite phenomenon takes place for matter.

There is no redshift transition from deceleration to acceleration, i.e., $z_{da}$ for $b^2 = 0.2$ and $\alpha=0.08$.
\begin{figure}[h!]
\begin{center}
\includegraphics[height=4in, width=7in]{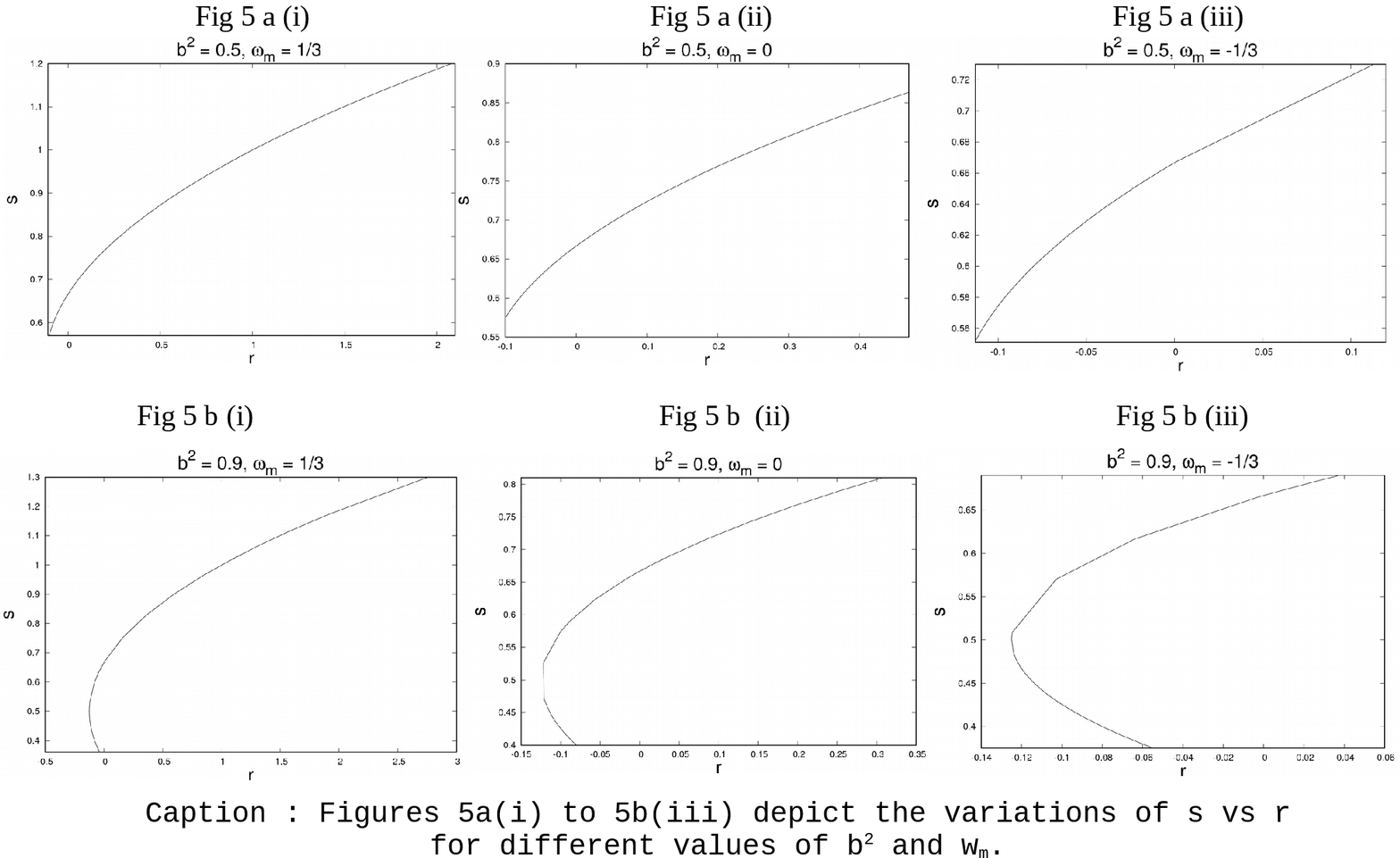}~~~
\end{center}
\end{figure}
Plots of $s$ vs $r$ for interacting cases of GCG are given in $5a(i)-5b(iii)$. For $b^2=0.9$, we find $s$ to be a double valued function of $r$, i.e., for a single $r$ we can have two $s$ values. For low $b^2$ there is only one branch where $\frac{ds}{dr}>0$ but for higher interaction, two branches with $\frac{ds}{dr}$ less than or equivalent to $ 0$ are found.  

The constraints on the statefinders of power-law cosmology from $H(z)$, $SNeIa$ and Big Bang Nucleosynthesis (BBN) observations will be studied now. The $H(z)$ data constrains the statefinders as $r = -0.09^{+0.04}_{-0.03}$ and $s = 0.58^{+0.04}_{-0.12}$ while the $SNeIa$ data constrains on statefinders as $r = -0.09^{+0.03}_{-0.02}$ and $s = 0.41^{+0.03}_{-0.03}$. The joint test of $H(z)$ and $SNeIa$ data puts the following constraints on statefinders : $r = -0.11^{+0.02}_{-0.01}$ and $s = 0.44^{+0.03}_{-0.03}$. The errors in the above values are at $1\sigma$ level. The primordial nucleosynthesis restricts $r ~greater ~than ~or~ approximately ~equal~ to~ 1.15 $. We observe that the best fit values of statefinders $r$ and $s$ predicted by observational $H(z)$ and SNeIa data are in conflict with the ones estimated by (BBN), as expected. 
\subsection{Modified Chaplygin Gas With Barotropic Fluid}
It has already been mentioned earlier that an interacting MCG model was considered in the literature \cite{COSMOLOGICAL_EVOLUTION_OF_MODIFIED_CHAPLYGIN_GAS}, nevertheless, analytic solutions for the continuity equations could not be determined assuming an interaction of the form $\Gamma = 3cH(\rho_d + \rho_m)$. Here, the authors of \cite{COSMOLOGICAL_EVOLUTION_OF_MODIFIED_CHAPLYGIN_GAS} performed a phase-space analysis and obtained a stable scaling solution at late times with the universe evolving into a phase of steady state. Since there is no microphysical hint on the nature of interactions between GCG and matter, we are bound to consider a phenomenological form of the interaction term.

From the equation (\ref{energy_density}) and the EoS of MCG we get, 
\begin{equation}\label{MCGpaper2continuity}
\int\frac{d\rho_d}{\rho_d  \left(1+A+b^2-\frac{B}{\rho_d ^{\alpha +1}}\right)} = ln\left(\frac{B'}{a^3}\right) \Rightarrow \rho_d = \frac{1}{(1+A+b^2)^{\frac{1}{\alpha +1}}}\left[B+ \left(\frac{B'}{a^3}\right)^{(1+A+b^2)(1+\alpha)}\right]^\frac{1}{1+\alpha}~~~~~~,
\end{equation}
where $A$ and $B$ are constants and $B'$ is the integrating constant.

Now, putting the above expression for $\rho_d$ in (\ref{Matter_density}) and multiplying both sides of the equation (\ref{MCGpaper2continuity}) by $a^3{(1+w_m)}$, the matter density $\rho_m$ can be evaluated as,

$$\rho_m = \int 3a^{2(1+ w_m)}\bigg[(1+w_m)- \Bigg\{\frac{( B+ (\frac{B'}{a^3})^{(1+A+b^2)(1+\alpha)} )}{(1+A+b^2)} \Bigg\}^{\frac{1}{(1+\alpha)}} da \bigg]=\frac{1}{a^{3(1+w_m)}} \times$$

$$ \left[C'' + a^{3(1+w_m)} b^2 \left\{1-\frac{B}{1+w_m}  \left(B+\bigg(\frac{B'}{a^3}\bigg)^{(1+A+b^2)(1+\alpha)}\right)^{\alpha }\left(1+\frac{1}{B}\bigg(\frac{B'}{a^3}\bigg)^{(1+A+b^2)(1+\alpha)}\right)^{-\alpha} \right.\right.$$ 
$$ \times \ (1+A+b^2)^{-\frac{1}{\alpha +1}} \times~ _2F_1\left[-\alpha ,-\frac{1+w_m}{(1+\alpha) (1+A+b^2)},1-\frac{1+w_m}{(1+\alpha) (1+A+b^2)},-\frac{1}{B}\bigg(\frac{B'}{a^3}\bigg)^{(1+A+b^2)(1+\alpha)}\right]$$
$$ +~~ \frac{(1+A+b^2)^{-\frac{1}{\alpha +1}}}{\alpha - w_m + (1+\alpha)(A+b^2)} \bigg(\frac{B'}{a^3}\bigg)^{(1+A+b^2)(1+\alpha)}\times \left(B+\bigg(\frac{B'}{a^3}\bigg)^{(1+A+b^2)(1+\alpha)}\right)^{\alpha } \left(1+\frac{1}{B}\bigg(\frac{B'}{a^3}\bigg)^{(1+A+b^2)(1+\alpha)}\right)^{-\alpha}$$ 
\begin{equation}
\left.\left.\ \times~ _2F_1\left[-\alpha ,\frac{\alpha - w_m + (1+\alpha)(A+b^2)}{(1+\alpha) (1+A+b^2)},\frac{\alpha - w_m + (1+\alpha)(A+b^2)}{(1+\alpha) (1+A+b^2)}+1,-\frac{1}{B}\bigg(\frac{B'}{a^3}\bigg)^{(1+A+b^2)(1+\alpha)}\right]\right\}\right]
\end{equation}
with the new integrating constant $C$. Note that for a prescribed matter EoS $w_m$, Our model consists of free parameters - the MCG parameter $A, B$, the coupling parameter $b^2$, the integrating constant $B'$ and $C''$. If one hampers, then $\Lambda$CDM has one free parameter($\Omega_{m0}$), while the most discussed dynamical DE model, $\Phi$CDM\footnote{The $\Phi$CDM model is a consistent dynamical dark energy model in which the currently accelerating cosmological expansion is powered by a scalar field $\Phi$ slowly rolling down an inverse power-law potential energy density.} has two free parameters ($\Omega_{m0}$ and $\alpha$)\cite{Peebles_Ratra, Cosmological_consequences}. In this interacting MCG model, the two integrating constants $B'$ and $C''$ can be fixed so that we shall also be left with only three free parameters, $A, B$ and $b^2$. Due to a high degree of nonlinearity in the expressions, it is very difficult to identify the relations of parameters with those occurring in the more well-known DE models. Now, the explicit expressions for the total energy density $\rho$ and the pressure $p$ can be written as \footnote{Mathematica Software was used to evaluate this integration.}

$$ \rho = \frac{1}{a^{3(1+w_m)}} \times \left[C'' + a^{3(1+w_m)} b^2 \left\{1-\frac{B}{1+w_m}  \left(B+\bigg(\frac{B'}{a^3}\bigg)^{(1+A+b^2)(1+\alpha)}\right)^{\alpha }\left(1+\frac{1}{B}\bigg(\frac{B'}{a^3}\bigg)^{(1+A+b^2)(1+\alpha)}\right)^{-\alpha} \right.\right.$$ 
$$ \times \ (1+A+b^2)^{-\frac{1}{\alpha +1}} \times~ _2F_1\left[-\alpha ,-\frac{1+w_m}{(1+\alpha) (1+A+b^2)},1-\frac{1+w_m}{(1+\alpha) (1+A+b^2)},-\frac{1}{B}\bigg(\frac{B'}{a^3}\bigg)^{(1+A+b^2)(1+\alpha)}\right]$$
$$+~~\frac{(1+A+b^2)^{-\frac{1}{1+\alpha}}}{\alpha - w_m + (1+\alpha)(A+b^2)} \bigg(\frac{B'}{a^3}\bigg)^{(1+A+b^2)(1+\alpha)}$$ 
$$\times \left(B+\bigg(\frac{B'}{a^3}\bigg)^{(1+A+b^2)(1+\alpha)}\right)^{\alpha } \left(1+\frac{1}{B}\bigg(\frac{B'}{a^3}\bigg)^{(1+A+b^2)(1+\alpha)}\right)^{-\alpha}$$

$$\left. \ \times~ _2F_1\left[-\alpha ,\frac{\alpha - w_m + (1+\alpha)(A+b^2)}{(1+\alpha) (1+A+b^2)},\frac{\alpha - w_m + (1+\alpha)(A+b^2)}{(1+\alpha) (1+A+b^2)}+1,-\frac{1}{B}\bigg(\frac{B'}{a^3}\bigg)^{(1+A+b^2)(1+\alpha)}\right]\Bigg\} \right]$$
\begin{equation}\label{density}
+~~\frac{1}{(1+A+b^2)^{\frac{1}{1+\alpha}}}\left[B+ \left(\frac{B'}{a^3}\right)^{(1+A+b^2)(1+\alpha)}\right]^\frac{1}{1+\alpha}~~and
\end{equation}
$$ p = \frac{w_m}{a^{3(1+w_m)}}\times \left[C'' + a^{3(1+w_m)}b^2\left\{1-\frac{B}{1+w_m}  \left(B+\bigg(\frac{B'}{a^3}\bigg)^{(1+A+b^2)(1+\alpha)}\right)^{\alpha }\left(1+\frac{1}{B}\bigg(\frac{B'}{a^3}\bigg)^{(1+A+b^2)(1+\alpha)}\right)^{-\alpha} \right.\right.$$ 
$$\times \ (1+A+b^2)^{-\frac{1}{1+\alpha}} \times~ _2F_1\left[-\alpha ,-\frac{1+w_m}{(1+\alpha) (1+A+b^2)},1-\frac{1+w_m}{(1+\alpha) (1+A+b^2)},-\frac{1}{B}\bigg(\frac{B'}{a^3}\bigg)^{(1+A+b^2)(1+\alpha)}\right]$$
$$+~~\frac{(1+A+b^2)^{-\frac{1}{1+\alpha}}}{\alpha - w_m + (1+\alpha)(A+b^2)} \bigg(\frac{B'}{a^3}\bigg)^{(1+A+b^2)(1+\alpha)}$$ 
$$\times \left(B+\bigg(\frac{B'}{a^3}\bigg)^{(1+A+b^2)(1+\alpha)}\right)^{\alpha } \left(1+\frac{1}{B}\bigg(\frac{B'}{a^3}\bigg)^{(1+A+b^2)(1+\alpha)}\right)^{-\alpha}$$

$$\left. \times~ _2F_1\left[-\alpha ,\frac{\alpha - w_m + (1+\alpha)(A+b^2)}{(1+\alpha) (1+A+b^2)},\frac{\alpha - w_m + (1+\alpha)(A+b^2)}{(1+\alpha) (1+A+b^2)}+1,-\frac{1}{B}\bigg(\frac{B'}{a^3}\bigg)^{(1+A+b^2)(1+\alpha)}\right]\Bigg\} \right]$$
\begin{equation}\label{pressure}
-~~\frac{A}{(1+A+b^2)^{\frac{1}{1+\alpha}}} \bigg[B+ \bigg( \frac{B'}{a^3} \bigg)^{(1+A+b^2)(1+\alpha)}\bigg]^\frac{1}{1+\alpha}-B \Bigg[(1+A+b^2)^{\frac{\alpha}{1+\alpha}}\Bigg\{B+ \bigg( \frac{B'}{a^3} \bigg)^{(1+A+b^2)(1+\alpha)}\Bigg\}^{-\frac{\alpha}{1+\alpha}}\Bigg]~~.
\end{equation}
The variations of the DE density $w_d = A - \frac{B}{\rho_d^{1 + \alpha}}$, the effective EoS parameter $w_{eff} = \frac{p}{\rho}$ and the deceleration parameter $q = \frac{3}{2}(1 + \frac{p}{\rho})-1$ for this interacting scenario can also be easily constructed using Eqs. (\ref{density}) and (\ref{pressure}). For further uncomplication we will use $B''$ instead of $B'^{(1+A+b^2)(1+\alpha)}$. We do not write them explicitly in order to avoid unnecessary expansion of the manuscript. Since the above expressions are quite complicated, it is very difficult to analyze the present model analytically. 

Instead, the variations of the relevant parameters, like $w_d$, $w_{eff}$ and $q$ against the redshift $z$ have been presented in figure 6, figures 7ai-7ciii and figures 8a-c repectively.
\begin{figure}[!ht]
\begin{center}
\includegraphics[height=9in, width=7in]{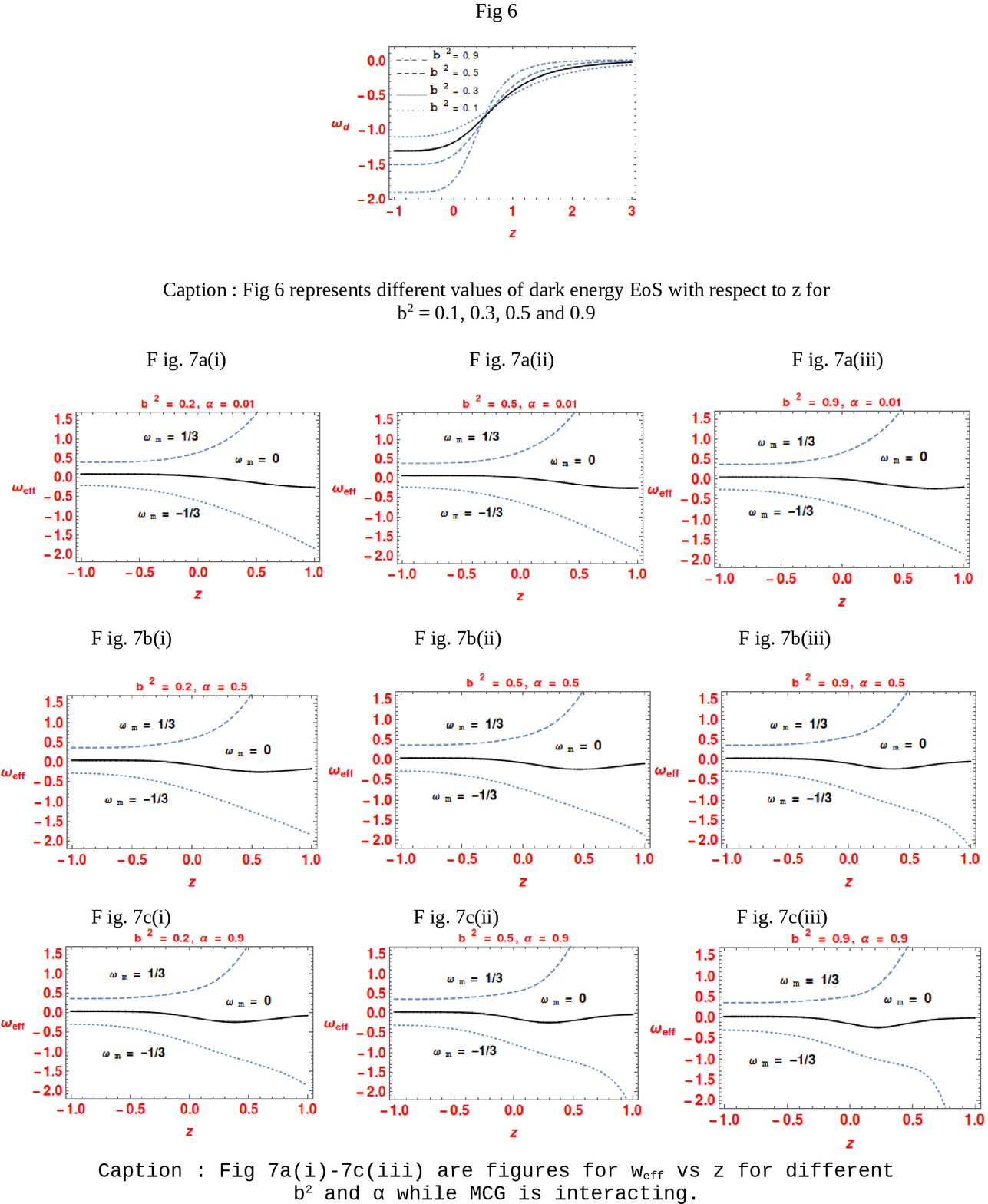}
\end{center}
\end{figure}
Fig. 6 depicts the variations of $w_d$ with respect to $z$. When $z$ is positive, i.e., when we are in past we see $w_d=0$ case to become asymptotic to $w_d = 0$ line or a line parallel to it. Though it has negative values all over there. As $z$ moves towards zero, $w_d$ decreases and in future, i.e., when $z$ is negative, the values of $w_d $ becomes asymptote with the $w_d = -1$ line or lines parallel to this. Physically, this can be interpreted as the EoS parameter for DE stays negative all over. But in future it becomes more negative. More the value of the interaction coefficient more will be the negativity of the DE EoS.

7ai to 7ciii depict the variations of $\omega_{eff}$ with respect to $z$. Contrary to the fact that for $\omega_m=0$, the graphs of  $\omega_{eff}$ was increasing for GCG case, we see the effective EoS to stay almost constant with increasing $z$. A region of z is found for which $\omega_{eff}$ is decreasing and then increasing and after that achieving a constant value again. As we increase $b^2$ and $\alpha$, this region shrinks. For $\omega=\frac{1}{3}$, the graph is increasing and $\omega_m=-\frac{1}{3}$ is decreasing always.

Deceleration parameter is a dimensionless measure of the cosmic acceleration. For $w_m = \frac{1}{3}$, i.e., for radiative matter mixed with DE we observe the deceleration parameter is positive in both past and future. This only becomes negative in the neighbourhood of $z = z_{{crit}_1}$ when $0 < z_{{crit}_1} < 0.5$ and $\left.\frac{\partial q}{\partial z}\right|_{z_{{crit}_1}}=0$. The same nature is valid for $w_m = 0$, i.e., for the pressureless dust case. But $q$ for $w_m = 0 <$ q for $w_m = \frac{1}{3}$ except the near neighbourhood of $z = z_{{crit}_1}$.
\begin{figure}[h!]
\begin{center}
\includegraphics[height=9in, width=7in]{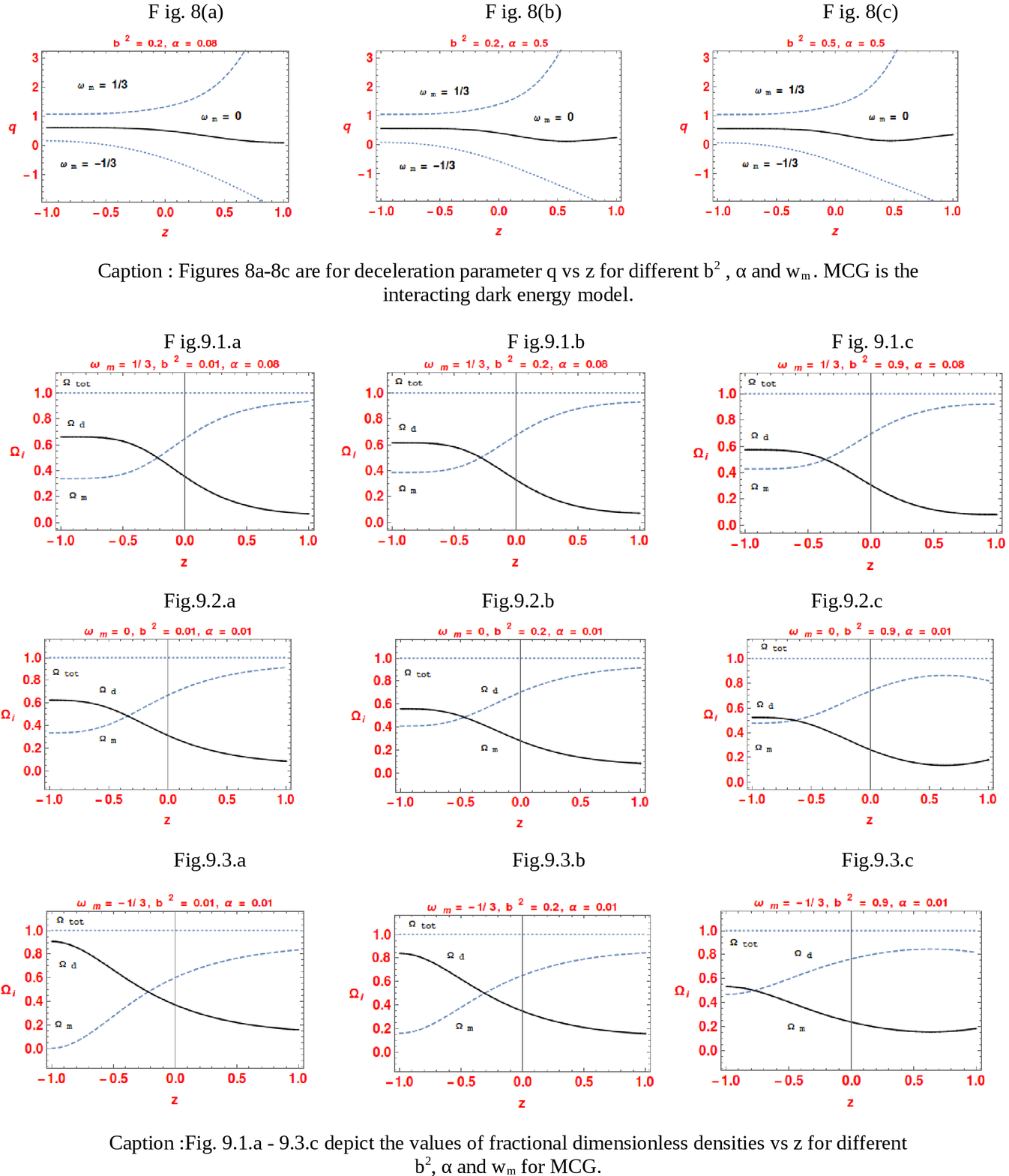}~~~~~
\end{center}
\end{figure}
For figure $9(b)$ and $9(c)$ the basic natures of the plots are almost same, but if $w_m$ is $0$ or $\frac{1}{3}$, we observe the transition line to have less inclination.

We have also calculated the redshift of transition from deceleration to acceleration ($z_{da}$) for $w_m = \frac{1}{3}$ and $0$. For $w_m = \frac{1}{3}$, we get $z_{da} = 0.06$ and $0.46$. For $w_m = 0$, we get $z_{da} = -0.06$ and $0.5$. There is no $z_{da}$ value for $w_m = -\frac{1}{3}$. We have constructed the table III for intracting MCG model for different values of MCG EoS parameter $\alpha$. Increment in $\alpha$ emforces the $z_{da}$ to shift towards more negative value, i.e., towards the future. Dimensionless density for DE increases for increment in $\alpha$.
\\
\textbf{TABLE III} : {\bf Values of the cosmological parameters with $A = 0.1,~ B = 0.01,~ B'' = 0.01, C'' = 1,~ w_m = -\frac{1}{3}. $}\\
\begin{center}
 \begin{tabular}{|| c | c | c | c ||}
 \hline
 Parameter & $b^2=0.2,~\alpha=0.08$ & $b^2=0.2,~\alpha=0.5$ & $b^2=0.5,~\alpha=0.5$ \\ [0.5ex]
 \hline
$z_{da}$ & -0.46281 & -0.6124510 & -0.639 \\
 \hline
  $q$ &  0.0000679138 & 0.0000909109 & 0.0000131338 \\
 \hline
 $\Omega_d$ & 0.0779723 & 0.146201 & 0.130128  \\
 \hline
 $\Omega_m$ & 0.922028 & 0.853799 & 0.869872  \\
  \hline
  $r$ & 0.000067923 & 0.0000909274 & 0.0000131342 \\
  \hline
  $s$ & 0.666712 & 0.6666727 & 0.666675 \\ [1ex]
  \hline
\end{tabular}
\end{center}

\begin{figure}[h!]
\begin{center}
\includegraphics[height=2in, width=7in]{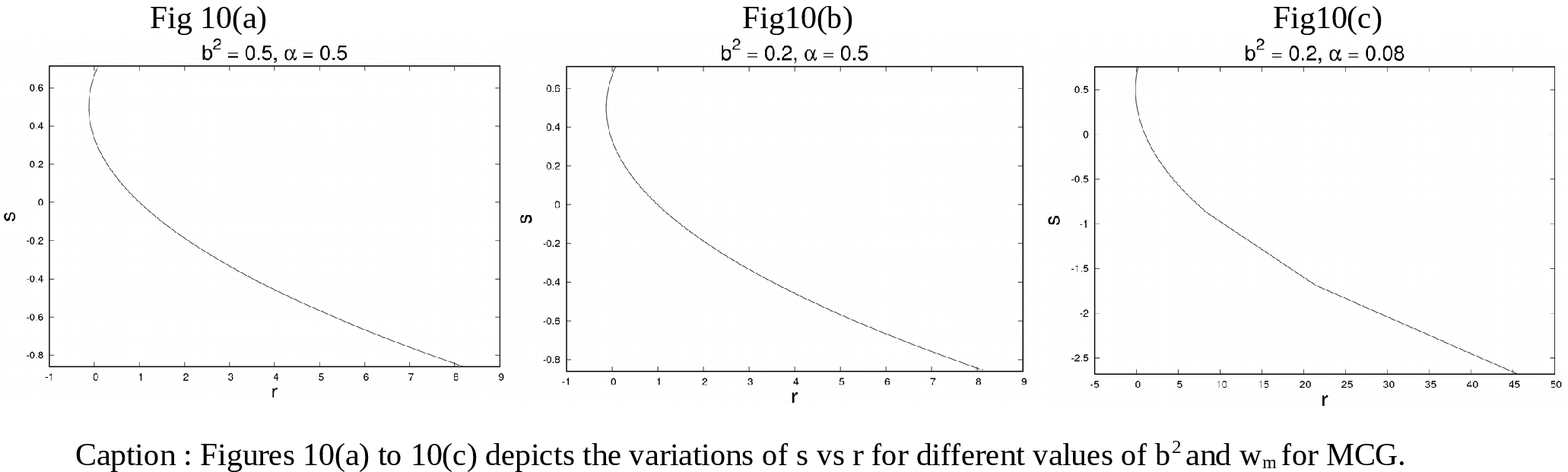}~~~
\end{center}
\end{figure}
In Fig. 10(a) we have plotted $s$ vs $r$ for MCG interacting with radiation. We observe that $r$ decreases with the increment of $s$ at first and then after reaching a local minima increases again. The rate of decrease and increase both are more if we consider the interaction of dust with MCG which has been shown in Fig. 10(b). Finally we draw Fig. 10(c) where interaction of MCG with fluid with EoS $w_m = -\frac{1}{3}$ is considered. We note that the decreasing $r$ vs $s$ branch dominates. The values of different thermodynamic parameters can be found in different literature \cite{Aviles} \cite{Capoziello} \cite{Mukherjee}  which are approximately equal or in the neighbouring range of the values found by us.
\section{Conclusions}
In this letter, we have considered the model of such fluids among which a particular type, popularly knowns as dark energy, is dominating all others. These fluids exert negative pressures which are theoretically responsible for the late time cosmic acceleration. In early epochs these fluids were not dominating and the expansion of universe was going through a decelerating phase. At a particular redshift we found it to be dominnt and to produce acceleration. This  points to the fact that either the quantity of dark energy has been increased with evolution of the universe, i.e., with cosmic time or the nature of it has been transited from a attracting nature to a repulsive nature. Assuming that the equation of continuity for the total matter content of the universe is intact, we have chosen the matter to get converted in dark energy. This leads to a interacting relation between normal matter and dark energy components of the universe. In this letter, the interaction term has been taken to be proportional with the Hubble's parameter as well as the dark energy density. A flat FLRW universe has been considered so far. We have constructed two different interacting dark energy plus matter models where the dark energy representative for two models are taken to be generalised and modified Chaplygin gas respectively. Through the interactive equation of continuity we have found the dark energy density as well as the matter density and hence the total density as a function of interacting coefficient $b^2$ and dark energy EoS parameters. It has been found that the dark energy EoS, $w_d$ turns to be negativee in the present time. For different $\alpha$, the values of $w_d$ are different in past but almost convergent to a same limit in present time. But for different $b^2$ the values of $w_d$ are distinct in future (through negative ). As high as $b^2$ is taken, i.e., as high as we take the rate of conversion from matter to dark energy, we see the dark energy EoS to become more negative infuture. In case of effecetive EoS for the Generalised Chaplygin gas interacting with barotropic fluid. We see the parameter is always decreasing with time for low $b^2$ and low $\alpha$ the rate of this decrease is slow if the latter fluid is at the quintessence barrier (i.e., $w_m = -\frac{1}{3}$) and high if the latter fluid acts like radiation (i.e., $w_m = \frac{1}{3}$). It physically says that there is no high tendency for matter to get converted into dark energy if it is already exerting negative pressure. On the contrary radiation like matter has a high tendency to get converted in dark energy depending on the value of the interaction coefficient. If interaction coefficient is enough high then the scope of matter to convert into dark energy increases. For MCG, however, the scenario is not the same. If the interacting matter is of  radiation nature, the effective EoS decreases abruptly with time in past and in future the rate of decrease becomes very slow. If the interacting matter shows pressureless dust type nature, it almost stays constant throughout (a downward hump in past arises). If the interacting matter has negative pressure, its $w_m$ increases (i.e., the negativity decreases) with time. So on comparision we can state that interacting GCG has positive effective EoS in past independent of the nature of the interacting matter. MCG does not follow the same pattern. 

While the deceleration parameter is studied, we see the curves are decreasing for GCG for almost all the values of $w_m$ and $b^2$. But for MCG, $q$ is decreasing with time only for $w_m = \frac{1}{3}.$ For $w_m = -\frac{1}{3}$ we observe it to increase in value.

For interacting GCG, dimensionless densities of dark energy and the matter form an horizonally directed structure taking $\Omega_i = 0$ and $1$ lines as the envelopes. The vertex of this hourglass slope moves from past to future as we increase $b^2$ we move from $w_m = \frac{1}{3}$ to $w_m = -\frac{1}{3}$ limits. Decrease in $w_m$  decreses the distances between $\Omega_d$ and $\Omega_m$ lines for past or future. For interacting MCG the basic structure stays the same. But the vertex of the hourglass is shifted towards future than the corresponding cases of GCG. Statefinder parameter $s$ for MCG is mainly decreasing with $r$ whereas that for GCG is increasing with $r$. We observe that the assumption of interaction helps to study the thermodynamic variables explicitely along with the nature of evolution of the universe. The values of redshifts for which deceleration to acceleration in universal expansion occurs have been found out for different values of the interaction coefficients and the dependencies are analysed. If no interactions were counted, the variation of $w_{eff}$ should linearly depend on $z$. We have verified the values of $z_{da}$ found and the corresponding $q(z_{da})$, $r(z_{da})$ and $s(z_{da})$-s with different observation based values found in previous works.

\vspace{.05 in}

{\bf Acknowledgment:}
This research is supported by the project grant of State Government of West Bengal, Department of Higher Education, Science and Technology and Biotechnology (File No.:- $ST/P/S\&T/16G-19/2017$). RB thanks IUCAA, Pune for Visiting Associateship.

RB dedicates this article to his PhD supervisor Prof. Subenoy Chakraborty as a tribute on his $60^{th}$ birth year.

Authors thank Mr. Aditya Kumar Naskar for a thoughtful reading of the manuscript.
  
\end{document}